\documentclass[fleqn,usenatbib,useAMS]{mnras}

\usepackage{graphicx}	
\usepackage{amsmath}	
\usepackage{multicol}        
\usepackage{bm}		
\usepackage{pdflscape}	

\defcitealias{Grayling_2024}{G24} 
\defcitealias{Thorp_2021}{T21}
\defcitealias{Grayling_2026b}{G26}

\usepackage{color}
\usepackage{orcidlink}

\usepackage[T1]{fontenc}
\usepackage{ae,aecompl}

\usepackage{newtxtext,newtxmath}

\title[A BayeSN view of the ZTF SN~Ia DR2]{On the origin of the environmental dependence of SN Ia magnitudes: A
BayeSN view of the ZTF SN Ia DR2}

\author[M. Ginolin et al.]{Madeleine Ginolin$^{1}$%
\thanks{Contact e-mail: \href{mailto:madeleine.ginolin@ast.cam.ac.uk}{madeleine.ginolin@ast.cam.ac.uk}}\orcidlink{0009-0004-5311-9301}, 
 Matthew Grayling$^{1}$\orcidlink{0000-0002-6741-983X}, 
 Kaisey S. Mandel$^{1,2}$\orcidlink{0000-0001-9846-4417}, 
 Maximilian Autenrieth$^{1,2}$\orcidlink{0009-0006-2068-5950}, 
\newauthor Benjamin M. Boyd$^{1}$\orcidlink{0000-0002-0622-1117},
Aaron Do$^{1}$\orcidlink{0000-0003-3429-7845}, 
 Lisa Kelsey$^{1}$\orcidlink{0000-0003-0313-0487}, 
 Matthew O’Callaghan$^{1}$\orcidlink{0009-0007-4567-7751}%
\\
$^{1}$Institute of Astronomy and Kavli Institute for Cosmology, University of Cambridge, Madingley Road, Cambridge CB3 0HA, UK\\
$^{2}$Statistical Laboratory, DPMMS, University of Cambridge, Wilberforce Road, Cambridge, CB3 0WB, UK}

\date{Last updated ---; in original form ---}

\pubyear{{\the\year{}}}

\begin{document}
\label{firstpage}
\pagerange{\pageref{firstpage}--\pageref{lastpage}}
\maketitle

\begin{abstract}
Astrophysical variabilities of Type Ia supernovae (SNe~Ia), such as their link with their birth environment, are now one of the leading sources of systematic uncertainties on the measurement of the dark energy equation-of-state parameter $w$. Population studies of SNe~Ia, using large samples, give precious insights into these variabilities. We analyse a volume-limited subsample of 932 SNe from the ZTF SN Ia DR2 with BayeSN, a hierarchical Bayesian model for SN Ia SEDs. We investigate the distributions of SN~Ia light curve parameters and their link with SN environment. Using a new training of BayeSN released in a companion paper, we find a smaller scatter of Hubble residuals compared to SALT. We then investigate the magnitude step, which accounts for the correlation between SN~Ia standardised absolute magnitude and host environments. We find a posteriori steps of $0.103\pm0.010$ mag (a $10.1\sigma$ difference from 0) when using global stellar mass as an environmental proxy, and $0.086\pm0.010$ mag ($8.3\sigma$) when using local colour, in accordance with steps computed using SALT light curve fits. This confirms that the large step seen in the ZTF SN~Ia DR2 data was not due to the SALT fit or the associated standardisation process. We then investigate the origin of the step, using a BayeSN model which accounts for both an intrinsic magnitude step and differing dust properties with the SN environment. We find a $0.103\pm0.018$ mag ($5.6\sigma$) step in global mass and a $0.085\pm0.019$ mag ($4.5\sigma$) step in local colour. The means of the $R_V$ distribution are similar between different host environments, with $\Delta\mathbb{E}(R_V)\leq0.2$ across all environment proxies, with significances ranging from $0.6\sigma$ to $1.2\sigma$. This is a strong signal of the existence of an intrinsic dependence of SN~Ia absolute magnitude on environment.
\end{abstract}




\newpage

\section{Introduction}

Type Ia supernovae (SNe~Ia) are one of the most used tools to measure distances in the Universe, as their peak absolute magnitudes are, at first order, a constant, making them \textit{standard} candles. They have been instrumental in the discovery of the acceleration of the expansion of the Universe \citep{Riess_1998, Perlmutter_1999}. They are still key in measurements of cosmological parameters, such as the Hubble constant $H_0$ \citep{Freedman_2021, Riess_2022} and the dark energy equation-of-state parameter $w$ \citep{DES5YR, Rubin_2023}. As such, they are at the heart of two puzzles of today's cosmology: the Hubble tension, a persistent mismatch of $H_0$ between local measurements and constraints from the Cosmic Microwave Background \citep[see e.g.][]{Di_Valentino_2025}, and the recent tentative detection of an evolution of $w$ with cosmic time \citep{DESI_2024, DESI_2025}. 

In reality, SNe~Ia are \textit{standardisable} candles: their peak absolute magnitudes exhibit a natural scatter of $\sim 0.4$ mag. In order to reduce this scatter, to increase the precision of distance measurements derived from SNe~Ia, an empirical standardisation was progressively developed. This process exploits correlation of the peak absolute magnitude with light curve properties e.g. its width \citep{Phillips_1993} and its colour \citep{Tripp_1998}, reducing this scatter to $\sim 0.15$ mag. An additional standardisation parameter was added in the early 2010s, which accounts for correlations between the host environment of the SN and its (standardised) absolute magnitude \citep{Kelly_2010, Sullivan_2010, Lampeitl_2010, Childress_2013}. Originally dubbed the mass step, as it was based on the stellar mass of the host galaxy of the SN, studies of this dependency have grown more numerous in recent years. For example, different tracers to measure this step have been investigated, and in particular tracers local to the SN, thought to measure more accurately its environment \citep{Rigault_2020, Briday_2022, Kelsey_2021}. While it is now usually included in cosmological analyses \citep{Betoule_2014, Brout_2022, DES5YR}, the incorrect or partial correction of its effect might bias cosmological parameter inference \citep{Rigault_2015}.
Moreover, the origin of this step is still debated. While some argue that it is linked to an intrinsic difference of SN~Ia, such as differences in the progenitor system, e.g. age, metallicity, explosion mechanisms \citep{Rigault_2020, Briday_2022, Kim_2018, Sarin_2026, Magee_2026}, another hypothesis has arose in recent years, attributing this step to different dust properties in different host environments \citep{Brout_Scolnic_2021, Popovic_2023}. In reality, it is likely that the environmental step is a mixture of both effects \citep{Wiseman_2022, Thorp_2021,Thorp_2022, Kelsey_2023, Popovic_2024b, Grayling_2024, Hayes_2025}.

SN~Ia properties also correlate with the astrophysical environment in which they exploded. The correlation between SN stretch, an intrinsic light curve property, and environment, where older and redder galaxies host faster evolving SNe on average \citep{Filippenko_1989, Hamuy_1996, Sullivan_2010, Rigault_2020}, has been known since the 90s. In recent years, in-depth studies of this correlation were published, along with the development of analytical models \citep{Nicolas_2021, Ginolin_2025b}. Most of these are based on a two-population SN~Ia model \citep{Mannucci_2006, Sullivan_2006, Smith_2012, Childress_2014, Rigault_2013}, similar to the intrinsic difference postulated as the origin of the step. 
On the other hand, SN light curve colour is thought to be a mixture of intrinsic SN colour variations due to the explosion physics of the SN and dust reddening \citep{Jha_2007, Mandel_2017, Uddin_2020, Duarte_2023, Wojtak_2023}, which will necessarily change with environment \citep{Brout_Scolnic_2021, Popovic_2023, Popovic_2025}. This is also translated through the expected variation of the colour standardisation parameter $\beta$ with host galaxies parameters, e.g. host mass \citep{Sullivan_2010, Amanullah_2010, Mandel_2017, Gonzalez_Gaitan_2021, Brout_Scolnic_2021, Rubin_2023, Rubin_2026}.

Most of the analyses cited in the previous paragraphs have been based on SALT light curve fits \citep{Guy_2005, Guy_2007, Betoule_2014, Kenworthy_2021}. While SALT is widely used in cosmological analyses, its model of SN~Ia SED is mostly empirical. In particular, SALT colour attempts to encapsulate both intrinsic and extrinsic colour within a single apparent colour parameter $c$, without leveraging our understanding of dust reddening. Correctly modelling for dust is crucial for a precise determination of cosmological parameters, as it is now one of the dominating systematics in the $w$ error budget \citep{Vincenzi_2024}. 
BayeSN \citep{Mandel_2022, Thorp_2021} solves this issue by explicitly including the impact of dust on the intrinsic SN~Ia SED. It can thus provide precious insights on the distribution of SNe properties and the associated dust properties, as well as on the origin of the environmental step, as was done in \cite{Thorp_2022, Grayling_2024}.

We here analyse the Zwicky Transient Facility (ZTF) sample with BayeSN. We use the same selection cuts as in \cite{Ginolin_2025a, Ginolin_2025b}, to be able to cross-check results with this SALT-based analysis. In particular, we use a volume-limited subsample of the DR2, which is designed to make non-random selection effects negligible.

We provide a brief description of BayeSN in Sec.~\ref{sec:bayesn}, as well as the particular model used in this analysis, described in  \cite{Grayling_2026b}, hereafter \citetalias{Grayling_2026b}. We present in Sec.~\ref{sec:data} the data used. In Sec.~\ref{sec:lc_fits}, we show BayeSN fits of the ZTF light curves. We investigate the two light curve properties, $A_V$ and $\theta_1$, as well as the Hubble residuals, and compare the results with the SALT-based analysis of \cite{Ginolin_2025a, Ginolin_2025b}. In Sec.~\ref{sec:step}, we focus on the magnitude step, using a BayeSN model which includes dust variations across environments. We discuss these results and analysis variations in Sec.~\ref{sec:discussion}. We conclude in Sec.~\ref{sec:conclusion}.

\section{BayeSN}
\label{sec:bayesn}

BayeSN is a hierarchical Bayesian model of SN~Ia at the SED level, which describes a population of SNe, and samples a posterior of both population-level and individual-supernova parameters. The BayeSN framework is described in \cite{Mandel_2022} (building upon previous hierarchical SN~Ia light curve models of \citealt{Mandel_2009,Mandel_2011}), and a first training was released in \cite{Thorp_2021}, hereafter \citetalias{Thorp_2021}. The current BayeSN code, based on the \texttt{numpyro} implementation described in \cite{Grayling_2024}, is publicly available and documented \footnote{\url{https://bayesn.readthedocs.io/}}. 
We provide in Sec.~\ref{subsec:bayesn_model} a brief general description of the BayeSN model, and introduce the \citetalias{Grayling_2026b} training, used in this analysis in Sec.~\ref{subsec:calibrayesn}.

\subsection{Description of the model}
\label{subsec:bayesn_model}

In the BayeSN model, the SED of a given SN, indexed as $s$, is written as:
\begin{multline}-2.5\log\left(\frac{S^s(t,\lambda)}{S_0(t,\lambda)}\right)=M_0+\delta M^s+W_0(t,\lambda)+\theta_1^s \, W_1(t,\lambda) \\
    +A_V^s \, \xi(\lambda; R_V^{(s)})+\epsilon^s(t,\lambda),
    \label{eq:bayesn}
\end{multline}
where $t$ is the rest-frame $B$-band phase, and $\lambda$ is the rest-frame wavelength. Here, and throughout the paper, $\log$ denotes the decimal logarithm. The base template $S_0(t,\lambda)$ is the \cite{Hsiao_2007} SED template combined with a normalisation constant $M_0$, fixed at $-19.5$ mag.
The rest of the components of the model are described below:
\begin{itemize}
    \item $\delta M^s$ is a per-SN achromatic offset, drawn from a normal distribution, such as $\delta M^s\sim N(0, \sigma_0^2)$. This term captures the unexplained scatter of peak absolute magnitude, parametrised by an intrinsic scatter $\sigma_0$.
    \item $W_0(t,\lambda)$ and $W_1(t,\lambda)$ are respectively the $0^\mathrm{th}$ and $1^\mathrm{st}$ order functional principal component (FPC) of the SED, implemented as cubic spline surfaces.
    \item $\theta_1^s$ is a per-SN coefficient that quantifies the impact of $W_1$ on the final SED. It is correlated to the width of the light curve, and as such the product of $\theta_1^s$ and $W_1$ reproduces the 'broader-brighter' relation \citep{Phillips_1993}, where slower evolving SNe, which have broader light curves at peak, are on average brighter than fast-evolving SNe. We use a Gaussian prior on $\theta_1$, so that $\theta_1\sim N(0,1)$.
    \item $\xi(\lambda; R_V^{(s)})$ is the dust extinction law, here from \cite{Fitzpatrick_1999}, parametrised by $R_V^{(s)}$. $R_V$ can either be treated as a population parameter, or be fitted individually for each SN. We use a fixed $R_V$ for Section~\ref{sec:lc_fits}, and we investigate the use of a per-SN $R_V^s$ in Sec.~\ref{sec:step}.
    \item $A_V^s$ is the per-SN dust extinction in V-band. It is assigned an exponential prior, parametrised by a population parameter $\tau_A$, so that 
    \begin{align*}
        P(A_V^s|\,\tau_A) &= H(A_V^s) \times \tau_A^{-1}\exp(-A_V^s/\tau_A)
        \\ &= \begin{cases}\tau_A^{-1}\exp(-A_V^s/\tau_A) \mathrm{~if~} A_V^s\geq0, \\ 0 \mathrm{~if~}A_V^s<0.\end{cases}
    \end{align*}
    where $H$ is the Heaviside function.
    \item $\epsilon^s(t,\lambda)$ is a per-SN term that captures the intrinsic colour variations unaccounted for by the dust term $A_V^s \, \xi(\lambda; R_V^{(s)})$. It is represented by a cubic spline over two dimensions, defined by a matrix of knots $\boldsymbol{E}^s$, drawn from a multivariate Gaussian $\boldsymbol{e}^s\sim N(0,\boldsymbol{\Sigma}_\epsilon)$. Here $\boldsymbol{e}^s$ is the vectorisation of $\boldsymbol{E}^s$, and $\boldsymbol{\Sigma}_\epsilon$ is a covariance matrix, inferred as a hyperparameter during the training stage of the model.
\end{itemize}
The resulting SED is then corrected for Milky-Way extinction, scaled to the distance modulus $\mu^s$, and can then be integrated through bandpasses to produce photometric points, which are then compared to measured light curve data. $\mu^s$ is fitted using a prior which combines redshift information and an assumed cosmology. This choice allows for magnitude information to be used as an additional constraint on dust properties when training, which is especially useful in cases where the wavelength coverage of the data is relatively narrow, such as in the ZTF dataset (see Sec.~\ref{sec:data}). The distance modulus is thus conditioned as $P(\mu^s|z^s)=N(\mu_\mathrm{cosmo}(z^s), (\sigma_\mu^s)^2)$, where $\mu_\mathrm{cosmo}$ is computed assuming a flat $\Lambda$CDM Universe with $\Omega_m=0.28$ and $H_0=73.24\mathrm{~km~s}^{-1}\mathrm{~Mpc^{-1}}$, and  $(\sigma_\mu^s)^2\approx(\frac{5}{z\ln 10})^2 [(\sigma_z^s)^2+\sigma_\mathrm{pec}^2/c^2$], where $\sigma_z^s$ is the redshift uncertainty and $\sigma_\mathrm{pec}=150\mathrm{~km~s}^{-1}$ is the peculiar velocity dispersion \citep{Carrick_2015}. While the assumption of a cosmology would need to be considered when using BayeSN on high redshift and magnitude-limited samples, this is not a concern here. Indeed, we use a low-redshift sample, for which the distance modulus is only marginally sensitive to values of cosmological parameters, in its volume-limited version (see Sec.~\ref{subsec:vollim}), essentially removing any selection bias.

BayeSN works in two modes: fitting and training. When fitting data with a pre-trained model, we only infer individual SN properties, denoted by the superscript $s$ ($A_V^s$, $\theta_1^s$, $\delta M^s$ and $\epsilon^s$). As we do not need to infer population-level dust properties ($R_V$ and $\tau_A$), distances are kept as a free parameter and no cosmology-dependent prior is used. We use the fitting mode in Sec.~\ref{sec:lc_fits}. In Sec.~\ref{sec:step}, we partially train the model to jointly infer per-SN properties and population-level dust parameters. We however keep $W_0$, $W_1$, $\sigma_0$ and $\boldsymbol{\Sigma}_\epsilon$ fixed to their model value. We note that in that section, we infer an individual $R_V^s$ for each SN. We discuss the full training of BayeSN with ZTF in Sec.~\ref{subsec:full_training}.

In each of the runs of this analysis, we infer the posteriors of the parameters using MCMC sampling. We use four chains initialised at different locations of the parameter space. Each chain is run for 500 steps, including 250 warm-up steps.  For every analysis, we evaluate the convergence of the chains, both by visually inspecting them (especially when inferring model hyperparameters), and by computing the potential scale reduction factor $\hat{R}$ for each parameters.

\subsection{BayeSN-Dovekie model}
\label{subsec:calibrayesn}

In our fiducial analysis, we use the soon-to-be released \citetalias{Grayling_2026b} BayeSN model (not to be confused with the recent extended-phase optical-NIR BayeSN model used for lensed SNe~Ia from \citealt{Grayling_2026}). This model was obtained using a framework that simultaneously trains a SED model in a BayeSN-like fashion, and cross-calibrates bandpasses across different surveys in a Dovekie-like fashion \citep{Dovekie}. Cross-calibration puts bandpasses in a common photometric system, and is necessary when using multiple surveys, and thus multiple observing systems. Dovekie builds on previous cross-calibration efforts \citep{Fragilistic}, and adds two main features: the estimation of the uncertainty on filter offsets and shifts, and a new calibration path using hierarchical Bayesian modelling of DA white dwarfs \citep{Boyd_2025}. The cross-calibration (filter offsets and shifts) is obtained using information from the SNe Ia themselves, combined with priors coming from previous Dovekie runs.  

\citetalias{Grayling_2026b} also used different data than the fiducial \citetalias{Thorp_2021} training. \citetalias{Thorp_2021} used 157 SNe from Foundation \cite{Foley_2018}, while \citetalias{Grayling_2026b} uses three low-redshift surveys (Foundation, CSP from \citealt{CSP}, CfA3-4 from \citealt{CfA3,CfA3bis,CfA4}), along with four higher-redshift surveys (PS1 from \citealt{PS1}, SDSS from \citealt{SDSS}, SNLS from \citealt{Astier_2006}, DES from \citealt{Abbott_2019}), totalling 1024 SNe. We evaluate the impact on our results of the BayeSN training used in Sec.~\ref{subsec:training_impact}.

\section{Data}
\label{sec:data}

In this paper, we use the second SN~Ia release of ZTF \citep{Bellm_2019, Graham_2019, Masci_2019, Dekany_2020}. ZTF is a low-redshift transient photometric survey, observing the northern sky. It uses three bands ($g$, $r$ and $i$), and can detect SNe~Ia up to $z\sim0.1$. It additionally comprises a dedicated low-resolution integral field spectrograph, the SEDmachine \citep{SEDm, IFU}, optimised for the spectroscopic typing of transients.

\subsection{ZTF SN~Ia DR2}

The ZTF SN~Ia DR2, described in details in \cite{Rigault_2025a}, is composed of all spectroscopically typed SNe~Ia observed by ZTF between March 2018 and December 2020. The spectroscopic typing was done mostly with spectra coming from the BTS survey \citep{Perley_2020}. The data made publicly available includes the light curves of the SNe, as well as their SALT2.4 fits \citep{Guy_2007, Guy_2010, Betoule_2014, Taylor_2021}.

Each SN also has an associated redshift, coming either from a spectrum of its host galaxy (61\%), from host galaxy features in the SN spectra (9\%) or from SN spectral features themselves (30\%).

Alongside with SN data, host galaxies and environmental properties were also released. The host matching was done using the $d_\mathrm{DLR}$ technique \citep{Sullivan_2006, Gupta_2016}, where $d_\mathrm{DLR}$ is the normalised distance of the SN to its host. The host galaxies were fitted with HostPhot \citep{HOSTPHOT} on Pan-STARRS \citep{Pan-STARRS} photometry. The derived properties are global stellar mass and rest-frame $(g-z)$ colour, and local stellar mass and $(g-z)$ colour, computed with an aperture of a 2 kpc radius around the SN. Host properties are obtained through SED fitting, following the method from \cite{Sullivan_2010, Smith_2020}.

The ZTF SN~Ia DR2 is not yet cosmology ready, due to known non-linearities and a photometric accuracy not yet sufficient for cosmological analysis \citep{Lacroix_2026}. We discuss the possible uses of the next release of ZTF, which fixes these issues, in Sec.~\ref{subsec:full_training}.

\subsection{Volume-limited sample}
\label{subsec:vollim}

We here use a volume-limited sample of the ZTF SN~Ia DR2, designed to minimise any observational biases, such as the Malmquist bias. We thus only use SNe up to $z=0.06$ \citep{Amenouche_2025}. We additionally use the same quality cuts as those described in \cite{Ginolin_2025a, Ginolin_2025b}, so that our samples are identical. These cuts are the following:
\begin{itemize}
    \item Light curve sampling: we keep SNe with seven detections (at the $5\sigma$ level) in the $[-10, 40]$ days phase range, in at least two bands and with at least two detections before peak and two detections after peak.
    \item SALT fit: we keep SNe with a light curve fit probability greater than $10^{-7}$. We also impose that the SALT stretch $x_1$ is in $[-3, 3]$, measured with a precision $\sigma_{x_1} <1$, that the SALT colour $c$ is in $[-0.2, 0.8]$ measured with a precision $\sigma_c < 0.1$. We finally impose that the error on the fitted time of peak luminosity $T_0$ is lower than one day.
    \item SN~Ia subtype: we remove peculiar SNe~Ia, such as 91bg or Ia-CSM, but keep the 91T.
\end{itemize}
Objects without host measurements (either local or global) are also removed from this sample.

The final sample comprises 932 objects. The list of every SN passing the cuts is available on GitHub \footnote{\url{http://github.com/mginolin/standax/blob/main/notebooks/Ginolin25ab_masterlist.csv}}.

We note that some objects were discarded from the sample on the basis of bad or extrapolated SALT fits, which might not be the case with BayeSN. We make this choice to allow for a meaningful comparison of the two fitting methods. 

\subsection{Preprocessing of the light curves}

In order to use the publicly available light curves, some preprocessing is needed. The first step is to output the light curves at a zero point of $\mathrm{ZP}=27.5$, to match the value expected by BayeSN. We then apply the quality cuts used in \cite{Rigault_2025a}, as implemented in the \texttt{lightcurve.get\_lcdata()} function from the \texttt{ztfcosmo} library. To speed up the fitting of the light curves and to limit memory usage, we use the same averaging recipe as the one used for the DR2 SALT fits, which performs averaging in 12h rolling bins. Finally, we apply the per-band error floors from \cite{Amenouche_2025}. We also compute CMB-frame redshifts, following \cite{Carr_2022}.

\section{Light curve fits}
\label{sec:lc_fits}

\begin{figure}
    \centering
    \includegraphics[width=1\columnwidth]{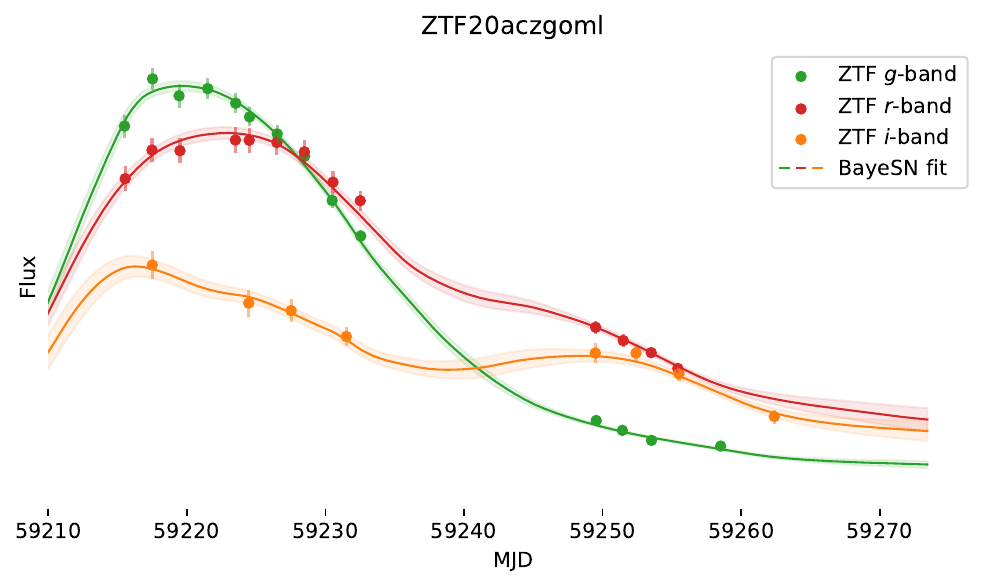}
    \caption{BayeSN fit of a randomly selected ZTF SN. The solid lines are the posterior mean of the model, while the shaded bands show the standard deviation of the chains.}
    \label{fig:example_good_fit}
\end{figure}

\begin{figure}
    \centering
    \includegraphics[width=1\columnwidth]{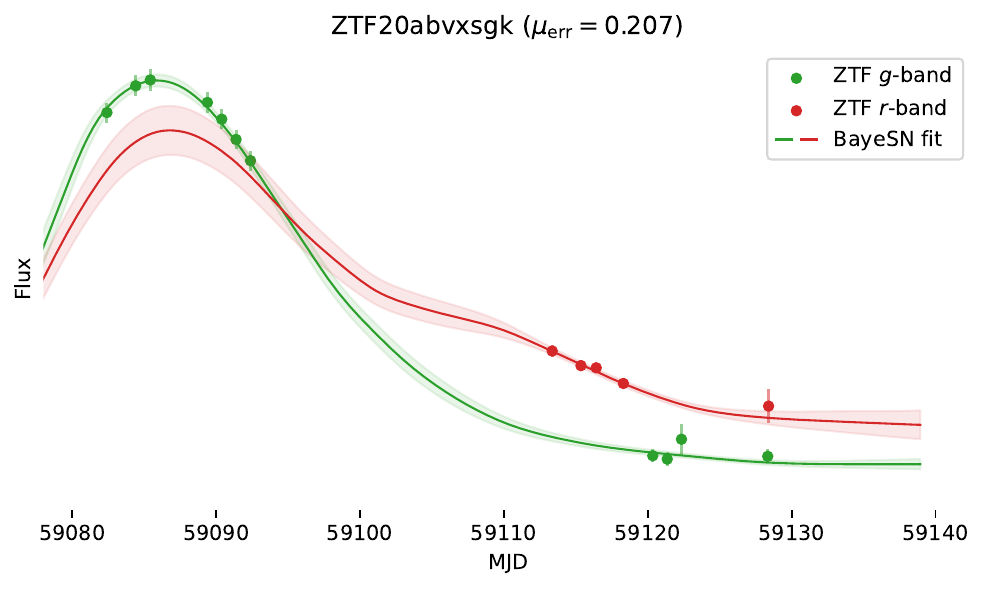}
    \caption{Example of a 'bad' BayeSN fit, defined as having an error on distance moduli larger than $\mu_\mathrm{err}>0.2$.}
    \label{fig:example_bad_fit}
\end{figure}

We start by fitting the DR2 volume-limited sample using the \citetalias{Grayling_2026b} BayeSN retraining presented in Sec.~\ref{subsec:calibrayesn}. The fitting of the 932 light curves takes around 30 mins when using four GPUs and a standard MCMC sampling. A variational inference (VI) implementation of the fitting is also available \citep{Uzsoy_2024}, but as our sample is of a reasonable size the speed-up enabled by VI is not necessary here. We present an example of a light curve fit in Fig.~\ref{fig:example_good_fit}. As expected, as we are dealing with an already selected sample, all of the fits are reasonably good, e.g. the MCMC chains are converged and well mixed. As an extra precaution, we manually check the few problem cases with large errors on the distance modulus ($\mu_\mathrm{err} > 0.2$). These seem to come from the lesser sampled light curves, or ones with only two bands ($g$ and $r$), as illustrated in Fig.~\ref{fig:example_bad_fit}. We thus conclude that we do not have to discard any extra SNe based on their BayeSN fit.

\subsection{Distribution of the parameters}
\label{subsec:param_distrib}

We now look at the distributions of the light curve parameters $\theta_1$ and $A_V$. In this section, to match the ZTF sample investigated in \cite{Ginolin_2025a, Ginolin_2025b}, we discard 7 additional SNe that did not pass the outlier rejection in \cite{Ginolin_2025a, Ginolin_2025b}, which was based on a cut on Chauvenet's criterion on the standardised SALT Hubble residuals.

\subsubsection{Light curve width $\theta_1$}
\label{subsubsec:theta_distrib}

\begin{figure}
    \centering
    \includegraphics[width=1\columnwidth]{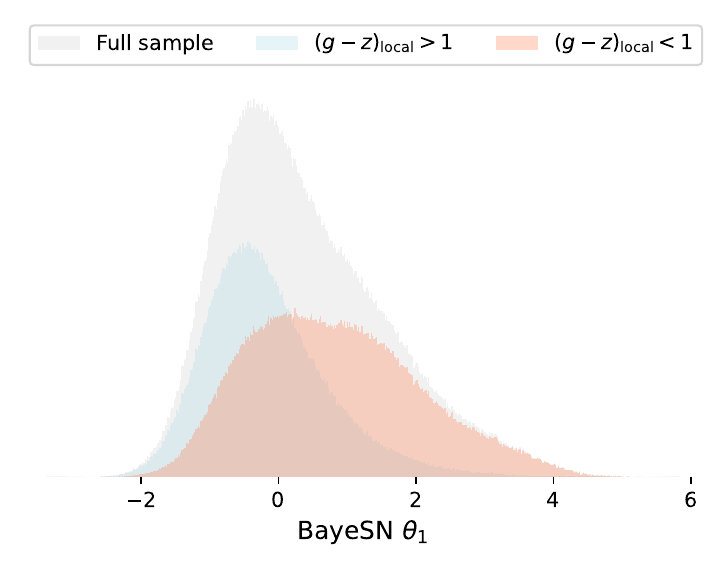}
    \caption{Histogram of BayeSN $\theta_1$. The full sample is plotted in grey, while the blue and red distribution correspond to SNe in locally blue and red environment, as defined in Sec.~\ref{subsubsec:theta_env}. To showcase both the population spread and the posterior uncertainty on $\theta_1$ for individual SNe, we use values from the full MCMC chains, minus warm-up, which results in 1,000 steps per SN.}
    \label{fig:theta_distrib}
\end{figure}

As seen in Fig.~\ref{fig:theta_distrib}, the distribution of $\theta_1$ for the full sample is non-Gaussian, with a second bump arising around $\theta_1\sim1$. This is similar to the behaviour of SALT stretch, and expected seeing their strong correlation (see Sec.~\ref{subsec:salt_comparison}). As $\theta_1$ is constrained by a Gaussian prior $\mathcal{N}(0,1)$, we manually look at the very large $\theta_1$ cases, in which it is far away from its prior. Those fits are reasonable, and correspond to very low SALT $x_1$ SNe. The existence of a large out-of-prior population might be due to a lower average stretch, corresponding to a higher average $\theta_1$, in the ZTF sample compared to the BayeSN training sample.

\subsubsection{Dust extinction $A_V$}
\label{subsubsec:av_distrib}

\begin{figure}
    \centering
    \includegraphics[width=1\columnwidth]{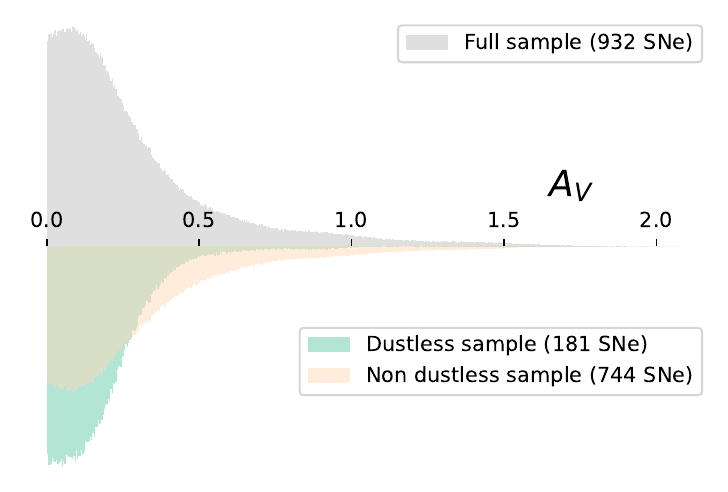}
    \caption{\textit{Top:} Histogram of BayeSN $A_V$. \textit{Bottom:} $A_V$ distributions for dustless and dusty subsamples, as defined in Sec.~\ref{subsubsec:av_env}.}
    \label{fig:av_vs_env}
\end{figure}

We plot on the top of Fig.~\ref{fig:av_vs_env} the distribution of $A_V$ for the full sample. Most of the SNe have an $A_V$ below 0.5, but the tails extends to value up to $A_V\sim1$. As we are using a volume-limited sample, it is expected that we observe redder SNe on average, i.e. larger $A_V$ values, that might be missing from magnitude-limited samples.

\subsection{Dependence on environment}
\label{subsec:env_dependency}

\subsubsection{$\theta_1$}
\label{subsubsec:theta_env}

We plot in Fig.~\ref{fig:theta_distrib} the $\theta_1$ distributions for locally blue and red environments, defined as $(g-z)_\mathrm{local} \lessgtr 1$. As expected, due to its link with the decline rate of the light curves, $\theta_1$ is strongly correlated with the host environment of the SN. In particular, SNe in locally red environments extend to larger values of $\theta_1$.

\subsubsection{$A_V$}
\label{subsubsec:av_env}

We then reproduce Fig.~1 from \cite{Ginolin_2025a}, where the colour distribution is split between a ``dustless'' sample and a non-dustless sample. The dustless sample is defined by objects in locally low stellar mass regions ($\log(M_\star/M_\odot) < 8.9$) and in the outskirts of their host galaxy ($d_\mathrm{DLR} > 1.2$). \footnote{This value differs from the one used in \cite{Ginolin_2025a}, as a small change in $d_\mathrm{DLR}$ values was not propagated through the dataset used in this analysis. We thus adapt the value of the cut to match the statistics of the dustless sample in \cite{Ginolin_2025a}.} The distance to the host galaxy is defined with the dimensionless Directional Light Ratio $d_\mathrm{DLR}$ \citep{Sullivan_2006, Gupta_2016}, a normalised measure of distance of the SN to the centre of the host, robust to different galaxy sizes and inclination effect. The non-dustless sample comprises of every SNe which are not in the dustless sample. Due to projection effects, SNe in this sample might also be unaffected by dust, which is why we do not name it the dusty sample. We see in the bottom of Fig.~\ref{fig:av_vs_env} that all high $A_V$ objects live in the non-dustless sample, as expected. However, as noted in \cite{Ginolin_2025a}, the dustless sample seems to still be affected by dust, as some of the $A_V$ values depart from 0. This is true if $A_V$ only encapsulates dust effect, and none of the intrinsic colour scatter present in the ZTF sample, as investigated in Sec.~\ref{subsec:salt_comparison}. The direct determination of $A_V$ made possible by BayeSN could lead to a more straightforward cut to isolate SNe relatively unaffected by dust, leading to a sample easier to standardise, with reduced systematic errors.

\subsection{SALT comparison}
\label{subsec:salt_comparison}

\begin{figure}
    \centering
    \includegraphics[width=1\columnwidth]{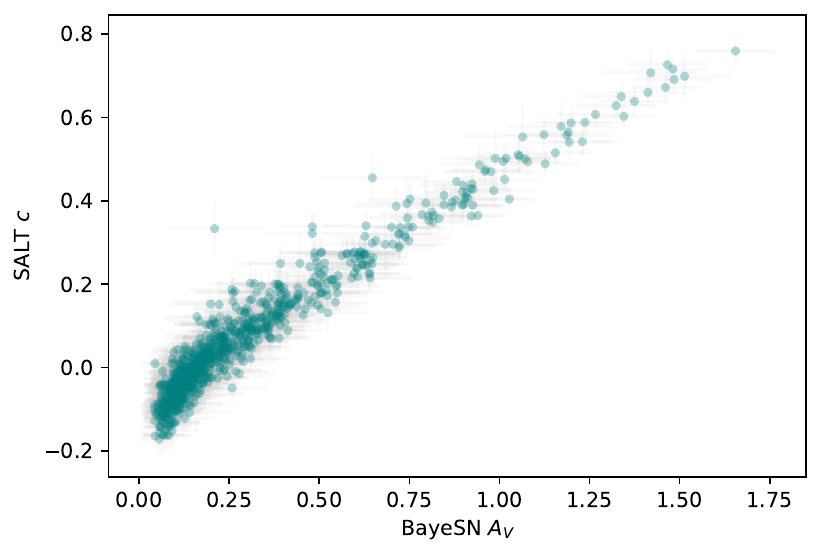}
    \caption{SALT colour $c$ against BayeSN extinction $A_V$.}
    \label{fig:c_vs_av}
\end{figure}

\begin{figure}
    \centering
    \includegraphics[width=1\columnwidth]{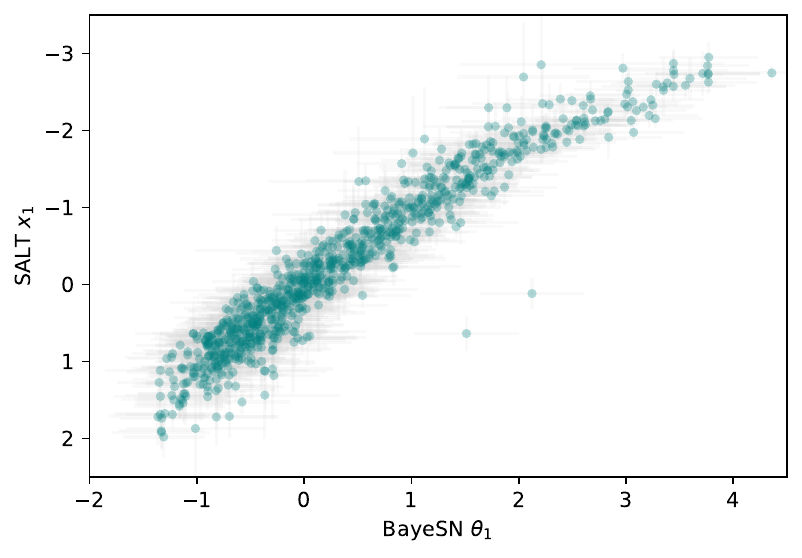}
    \caption{SALT stretch $x_1$ against BayeSN $\theta_1$. Note that the $x_1$ axis is reversed.}
    \label{fig:x1_vs_theta}
\end{figure}

We plot in Figs.~\ref{fig:c_vs_av}-\ref{fig:x1_vs_theta} the comparison between the BayeSN parameters ($\theta_1$ and $A_V$) and their SALT equivalents ($x_1$ and $c$). While these parameters are not a one-to-one match, this comparison is still interesting, as most of the literature on SN light curve properties, standardisation and their link with environment is based on SALT fits.

Looking at Fig.~\ref{fig:c_vs_av}, BayeSN $A_V$ and $c$ look strongly correlated. However, even at a null $A_V$, we see a scatter in $c$, which might point to a non-null intrinsic colour scatter. Fitting a Gaussian on the SALT $c$ distribution for SNe with $A_V<0.1$, we find a standard deviation of $0.027\pm0.004$ and a mean of $-0.083\pm0.004$. This is in accordance with the Gaussian intrinsic colour distribution fitted in \cite{Ginolin_2025a}, of width $0.030\pm0.005$ and mean $-0.085\pm0.004 $.

There is also a strong inverse correlation between $\theta_1$ and $x_1$, which however seems non-linear, as visible in Fig.~\ref{fig:x1_vs_theta}. The non-linearity of this relation is partly due to the fact that $\theta_1$ is a linear effect on log-flux, while $x_1$ is a linear effect on flux. It also might be a consequence of the non-linearity of the SALT stretch-residuals relation found in \cite{Ginolin_2025b}, where the Hubble residuals for low-stretch (i.e. high $\theta_1$) SNe have a stronger correlation with stretch (i.e. a larger $\alpha$). We investigate this question in more detail in Sec.~\ref{subsec:broken_theta}.

\subsection{Hubble residuals}
\label{subsec:hubble_res}

\begin{figure}
    \centering
    \includegraphics[width=1\columnwidth]{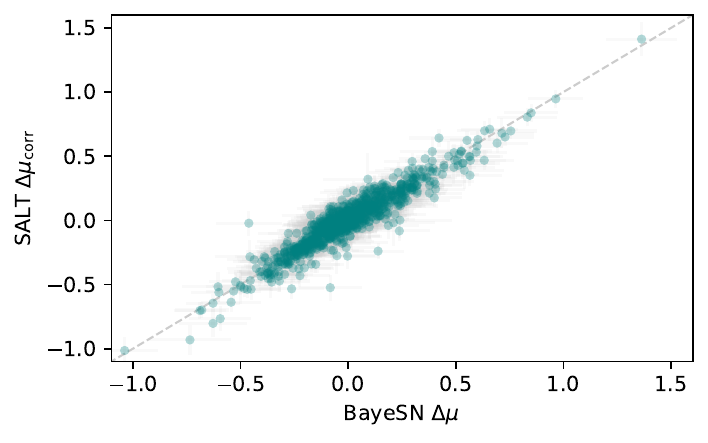}
    \caption{Standardised SALT Hubble residuals against BayeSN Hubble residuals, as described in Sec.~\ref{subsec:hubble_res}.}
    \label{fig:hubble_res}
\end{figure}

BayeSN Hubble residuals are defined as $\Delta\mu=\mu_\mathrm{fit}-\mu_\mathrm{cosmo}$, where $\mu_\mathrm{cosmo}$ are computed using the fiducial BayeSN cosmology (Flat $\Lambda$CDM with $H_0=73.24$ and $\Omega_m=0.28$, computed with \texttt{astropy} \citealt{astropy2013, astropy2018}). We subtract the mean of the Hubble residuals so that they are centred around zero. We compute SALT Hubble residuals using the \texttt{standax} \footnote{\url{https://github.com/mginolin/standax}} package. For the sake of comparison, we only use stretch and colour standardisation, and we do not include the environmental step correction, as this is not taken into account in the model used in this section. We discuss the inclusion of environmental dependencies in BayeSN in Sec.~\ref{sec:step}. We also do not include the broken-$\alpha$ found in \cite{Ginolin_2025b}. We use CMB-frame redshifts for both analyses. 
The comparison of Hubble residuals between SALT and BayeSN is plotted in Fig.~\ref{fig:hubble_res}. They are overall in good agreement. Some SNe~Ia appear to have large Hubble residuals, but consistently in both methods, pointing towards either an outlier such as a peculiar subtype of SN~Ia, or towards an issue with the light curve data itself.

\begin{figure}
    \centering
    \includegraphics[width=1\columnwidth]{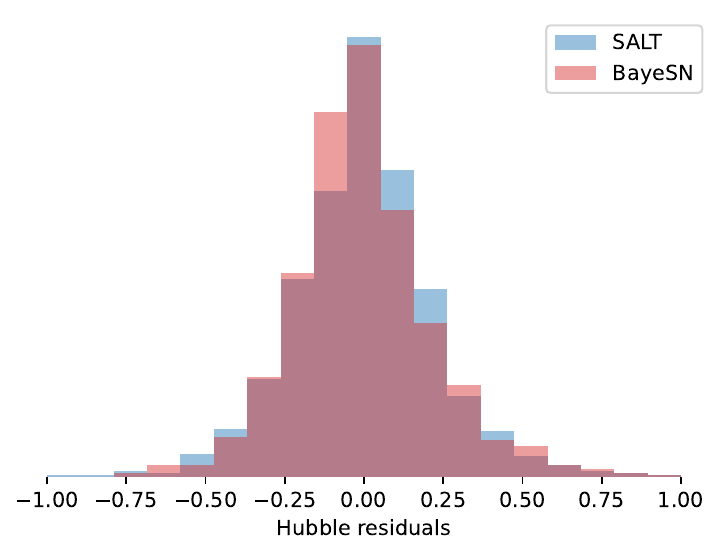}
    \caption{Histogram of (standardised) Hubble residuals, as computed in Sec.~\ref{subsec:hubble_res}, for SALT and BayeSN.}
    \label{fig:hubble_res_distrib}
\end{figure}

\begin{table}
 \caption{Hubble residuals scatter for BayeSN output $\mu_\mathrm{res}$ and standardised SALT outputs. We use both the regular standard deviation $\sigma_\mathrm{STD}$ and the normalised median absolute deviation $\sigma_\mathrm{nMAD}$. The host-$z$ subsample (729/929 SNe) is defined as objects with a redshift from host galaxy features, either from galaxy lines in the SN spectrum of from a spectrum of the host galaxy itself.}
 \label{tab:hubble_res_scatter}
 \begin{tabular}{lcccc}
  \hline
  Scatter & \multicolumn{2}{c}{Full sample} & \multicolumn{2}{c}{Host-$z$ sample} \\
   & BayeSN & SALT & BayeSN & SALT\\
   \hline
   $\sigma_\mathrm{STD}$ & 0.225 & 0.230 & 0.204 & 0.213 \\
  $\sigma_\mathrm{nMAD}$ & 0.177 & 0.181 & 0.165 & 0.169 \\
  \hline
 \end{tabular}
\end{table}

We now focus on the Hubble residual scatter. The distribution of Hubble residuals, both for BayeSN and SALT, is plotted in Fig.~\ref{fig:hubble_res_distrib}. We present in Table~\ref{tab:hubble_res_scatter} the value of the standard deviation $\sigma_\mathrm{STD}$ and the normalised Median Absolute Deviation $\sigma_\mathrm{nMAD}$, as it is more robust to outliers. We compute those scatter both for the full sample, and for a subsample of SNe~Ia for which the redshift comes from spectroscopic features of the host galaxy of the SN. We expect the scatter to be smaller for the host-$z$ sample, as the errors on $z$ are of the order of $\sigma_z\sim10^{-4}$, compared to the typical $\sigma_z\sim10^{-3}$ for redshifts coming from SNe features. 

The scatter is slightly smaller for BayeSN residuals, compared to SALT residuals, both for the STD and nMAD, and for both the full sample and the host-$z$ sample. \citetalias{Grayling_2026b} come to a similar conclusion, as they find a smaller scatter than SALT on their training sample, as well as an independent sample (DES5YR, \citealt{DES5YR-Dovekie}). This confirms that the \citetalias{Grayling_2026b} BayeSN training performs well on a sample on which it was not trained.

\subsection{A posteriori steps}
\label{subsec:steps_aposteriori}

\begin{figure}
    \centering
    \includegraphics[width=1\columnwidth]{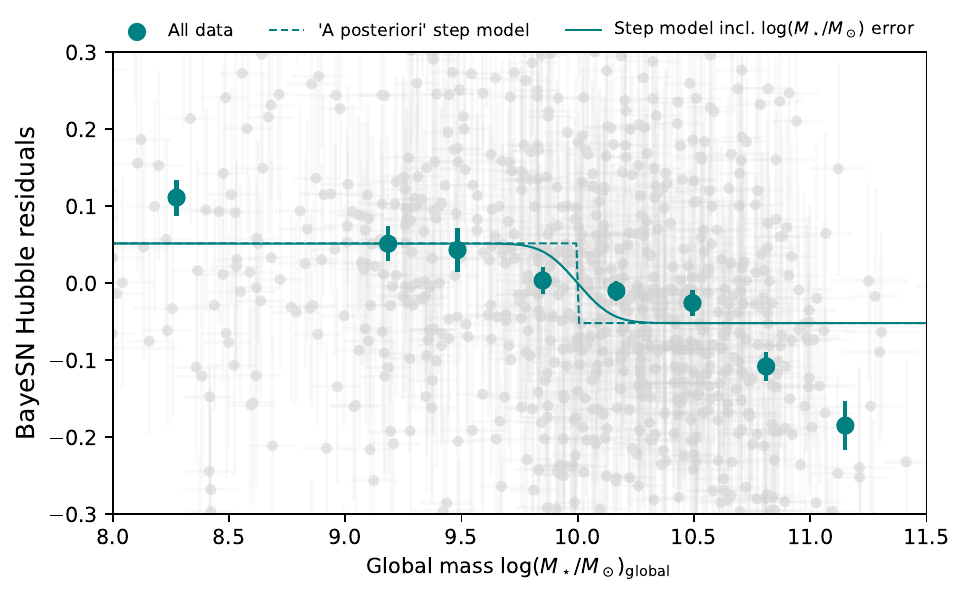}
    \caption{Hubble residuals for BayeSN light curve fits, as computed in Sec.~\ref{subsec:hubble_res}, against host global stellar mass. The blue points are the mean residuals per bins of global mass. The bins are equally spaced, except for the edge ones that are slightly larger to encompass all points. The 'a posteriori' step model (see Sec.~\ref{subsec:steps_aposteriori}) is illustrated in the blue dashed line, while the full line shows the step model when convoluted with the mean error on $\log(M_\star/M_\odot)$, as we account for these errors when computing the step.}
    \label{fig:step_gm}
\end{figure}

\begin{table}
 \caption{Magnitude steps (in mag) and their significances for the four DR2 environmental proxies. SALT steps are computed after standardisation, for a fair comparison, and are thus smaller than fiducial steps (as computed in \citealt{Ginolin_2025b}, see Sec.~\ref{subsec:steps_aposteriori} for more details).}
 \label{tab:bayesn_vs_salt_steps}
 \begin{tabular}{lcc}
  \hline
  Step tracer & BayeSN & SALT \\
   \hline
   $(g-z)_\mathrm{local}$ & $0.086\pm0.010$ ($8.3\sigma$) & $0.089\pm0.010$ ($8.5\sigma$)\\
   $(g-z)_\mathrm{global}$ & $0.080\pm0.010$ ($7.8\sigma$) & $0.083\pm0.010$ ($7.9\sigma$)\\
   $\log(M_\star/M_\odot)_\mathrm{local}$ & $0.070\pm0.010$ ($6.8\sigma$) & $0.087\pm0.011$ ($8.2\sigma$)\\
   $\log(M_\star/M_\odot)_\mathrm{global}$ & $0.103\pm0.010$ ($10.1\sigma$) & $0.099\pm0.010$ ($9.5\sigma$)\\
  \hline
 \end{tabular}
\end{table}

We here compute the leftover step in the BayeSN Hubble residuals, defined as the difference in Hubble residuals between two different SN environments, e.g. high- and low-mass hosts. This would be comparable to the a posteriori step (i.e. fitted after the SALT standardisation process) in \cite{Ginolin_2025b}. We use the same threshold value as in \cite{Ginolin_2025a, Ginolin_2025b}:
\begin{itemize}
    \item Global mass: $\log(M_\star/M_\odot)_\mathrm{global}^\mathrm{cut}=10$.
    \item Local mass: $\log(M_\star/M_\odot)_\mathrm{local}^\mathrm{cut}=8.9$.
    \item Local and global colour: $(g-z)^\mathrm{cut}=1$.
\end{itemize}
The proportion of SNe in each subsample defined by these cuts is roughly half-half ($54\%-46\%$ of SNe in locally red/blue environments, $57\%-43\%$ for globally red/blue environments, $52\%-48\%$ for locally high/low-mass environments, $53\%-47\%$ for globally high/low-mass environments).
We note that this method tends to underestimate the step compared to a full joint fit in the SALT framework, as some of the effect of the step is absorbed by the standardisation parameters ($\alpha$ and marginally $\beta$), and should thus not be used for standardisation purposes. It is however the only comparison currently available when using BayeSN in its fitting mode.
The amplitude of the steps are compiled in Table~\ref{tab:bayesn_vs_salt_steps}, and we illustrate the global mass step in Fig.~\ref{fig:step_gm}. We find consistent values within errorbars between SALT and BayeSN. The significance of the steps ranges between $\sim7-10\sigma$, for both SALT and BayeSN. This confirms that the large steps found in \cite{Ginolin_2025b} were not an artifact of SALT or of the SALT-based standardisation. We investigate more thoroughly the environmental steps in the next section.

\section{Investigating the environmental step}
\label{sec:step}

In this section, we leverage the explicit inclusion of dust impact on SN~Ia light curves in BayeSN to investigate the relative contribution of intrinsic SED differences and dust to the observed environmental magnitude step.

\subsection{Method}
\label{subsec:step_method}

To investigate the step, we build on the analysis of \cite{Grayling_2024}, hereafter \citetalias{Grayling_2024}. This analysis splits the SN sample into two based on the value of the stellar mass of their host galaxy, to investigate differences in their absolute magnitude (i.e. the mass step) and the dust between the two subsamples, by introducing and fitting for population-level parameters to capture these effects. This method was tested on SALT-based simulations, which ensured it could correctly disentangle dust effects from an intrinsic step \citep{Grayling_Popovic_2025}. In this Section, contrary to Sec.~\ref{sec:lc_fits}, an individual $R_V^s$ is inferred for each SN. To obtain reasonable constraints on dust parameters, we thus condition distances on redshift and a cosmology (see Sec.~\ref{subsec:bayesn_model}).  In \citetalias{Grayling_2024}, three different versions of the step training were presented. We here use the version in which the dust parameters ($R_V^s$ and $A_V^s$, see Sec.~\ref{sec:bayesn}) are drawn from two different populations, and an absolute magnitude offset is inferred between the two subsamples. We however extend the work of \citetalias{Grayling_2024} on two points. First, we use the four environmental tracers available in the ZTF SN~Ia DR2.
The second change made is the introduction of a smooth step. 

We illustrate this change using the global stellar mass $M_\star$ as the environmental tracer.
In \citetalias{Grayling_2024}, the data is split sharply into two according to the host mass. The host galaxy $R_V$ distribution is modelled as :
\begin{equation}
    R_V^s\sim \begin{cases} \mathrm{TN}(\mu_{R,\mathrm{LM}}, \sigma^2_{R,\mathrm{LM}}, 1.2, \infty) & \mbox{if } \log(M_\star^s/M_\odot) \leq 10 \\ \mathrm{TN}(\mu_{R,\mathrm{HM}}, \sigma^2_{R,\mathrm{HM}}, 1.2, \infty)  & \mbox{if } \log(M_\star^s/M_\odot) > 10
    \end{cases},
\end{equation}
where $\mathrm{TN}$ denotes a truncated normal distribution, and the subscript $_\mathrm{LM/HM}$ is used to distinguish parameters for the low- and high-mass sample .
Similarly, the achromatic offset is drawn from:
\begin{equation}
    \delta M^s\sim \begin{cases} N(\Delta M_0^\mathrm{LM}, \sigma^2_{0,\mathrm{LM}}) & \mbox{if } \log(M_\star^s/M_\odot) \leq 10 \\ N(\Delta M_0^\mathrm{HM}, \sigma^2_{0,\mathrm{HM}})  & \mbox{if } \log(M_\star^s/M_\odot) > 10
    \end{cases},
\end{equation}
and the extinction is drawn from:
\begin{multline}
    P(A_V^s|\,\tau_{A,\mathrm{~LM}},\tau_{A,\mathrm{~HM}}) = \begin{cases} H(A_V^s) \times \tau_{A,\mathrm{~LM}}^{-1}\exp(-A_V^s/\tau_{A,\mathrm{~LM}})\\ \mbox{~~~~~~~~if } \log(M_\star^s/M_\odot) \leq 10 \\ H(A_V^s) \times \tau_{A,\mathrm{~HM}}^{-1}\exp(-A_V^s/\tau_{A,\mathrm{~HM}})\\ \mbox{~~~~~~~~if } \log(M_\star^s/M_\odot) > 10
    \end{cases}.
\end{multline}

However, this formulation does not take into account the measurement errors on the mass. Indeed, for an object near the split value, and with a noisy mass measurement, there is comparable chances that it is on either side of the mass split. We thus adapt the smooth step used in \cite{Ginolin_2025b}. For example, when using global mass as our environmental proxy, we compute for each SN a parameter $p^s=\int_{-\infty}^{10}\mathcal{N}(x |\, m^s, (\sigma^s_m)^2)\, \rm{d}x$ where $m^s=\log(M^s_\star/M_\odot)$. For each parameter, the value for a given SN is drawn for a mixture of the high- and low-mass distributions. In the case of $R_V^s$, this is modelled as as: 
\begin{multline}
    R_V^s\sim  p^s\mathrm{TN}(\mu_{R,\mathrm{LM}}, \sigma^2_{R,\mathrm{LM}}, 1.2, \infty) \\ +(1-p^s) \mathrm{TN}(\mu_{R,\mathrm{HM}}, \sigma^2_{R,\mathrm{HM}}, 1.2, \infty).
\end{multline}
A similar transformation is applied to the $\delta M^s$ and $A_V^s$ distributions.

We thus fit for ten hyperparameters: $\Delta M_{0,\mathrm{~HM/LM}}$, $\sigma_{0,\mathrm{~HM/LM}}$, $\mu_{R,\mathrm{~HM/LM}}$, $\sigma_{R,\mathrm{~HM/LM}}$ and $\tau_{A,\mathrm{~HM/LM}}$.. We adopt the same hyperpriors as in \citetalias{Grayling_2024}. We proceed analogously when using the other three environmental tracers.

The impact of this smooth step is discussed in Sec.~\ref{subsec:smooth_step_impact}, and its implementation is available on GitHub\footnote{\href{https://github.com/mginolin/bayesn/tree/smooth_step}{\texttt{github.com/mginolin/bayesn/tree/smooth\_step}}}.

\subsection{Dust parameters}
\label{subsec:step_dust_params}

\begin{table*}
 \caption{Dust parameters split on environmental proxy (local and global colour, local and global mass). We also provide $\mathbb{E}(R_V)$ and $\sqrt{\mathrm{Var}(R_V)}$ as they do not correspond to $\mu_R$ and $\sigma_R$, since $R_V$ is drawn from a truncated Gaussian distribution.}
 \label{tab:dust_params}
 \begin{tabular}{lccccc}
  \hline
  Sample & $\tau_A$ & $\mu_R$ & $\sigma_R$ & $\mathbb{E}(R_V)$ & $\sqrt{\mathrm{Var}(R_V)}$ \\
  \hline
  \hline
   $(g-z)_\mathrm{local} > 1$ & $0.389\pm0.020$ & $2.145\pm0.111$ & $0.540\pm0.108$ & $2.206\pm0.076$ & $0.488\pm0.068$ \\
  $(g-z)_\mathrm{local} < 1$ & $0.341\pm0.020$ & $2.325\pm0.121$ & $0.577\pm0.109$ & $2.371\pm0.101$ & $0.535\pm0.077$ \\
  \hline
   $(g-z)_\mathrm{global} > 1$ & $0.394\pm0.021$ & $2.131\pm0.134$ & $0.551\pm0.123$ & $2.203\pm0.080$ & $0.492\pm0.074$ \\
  $(g-z)_\mathrm{global} < 1$ & $0.348\pm0.021$ & $2.296\pm0.139$ & $0.603\pm0.123$ & $2.358\pm0.103$ & $0.550\pm0.082$ \\
  \hline
   $\log(M_\star/M_\odot)_\mathrm{local} > 8.9$ & $0.475\pm0.024$ & $2.220\pm0.092$ & $0.527\pm0.088$ & $2.261\pm0.075$ & $0.488\pm0.062$\\
  $\log(M_\star/M_\odot)_\mathrm{local} < 8.9$ & $0.272\pm0.018$ & $2.274\pm0.175$ & $0.610\pm0.136$ & $2.347\pm0.124$ & $0.550\pm0.086$\\
   \hline
   $\log(M_\star/M_\odot)_\mathrm{global} > 10$ & $0.395\pm0.021$ & $2.164\pm0.110$ & $0.500\pm0.100$ & $2.207\pm0.083$ & $0.461\pm0.069$\\
  $\log(M_\star/M_\odot)_\mathrm{global} < 10$ & $0.359\pm0.021$ & $2.249\pm0.149$ & $0.681\pm0.129$ & $2.352\pm0.100$ & $0.597\pm0.078$\\
  \hline
 \end{tabular}
\end{table*}

We infer three dust parameters at the population level, for each of the two subsamples selected on environmental tracer: the mean and standard deviation of the $R_V$ distribution $\mu_{R,~\mathrm{HM/LM}}$ and $\sigma_{R,~\mathrm{HM/LM}}$, and the slope of the $A_V$ distribution $\tau_{A,~\mathrm{HM/LM}}$.

The values of each parameters are shown in Table~\ref{tab:dust_params}, and a summary of the differences between dust properties is shown in Table~\ref{tab:steps_dust} for all four environmental tracers. We do not detect any significant differences in $R_V$. The differences in mean $R_V$ are similar to the ones from \citetalias{Grayling_2024}, and smaller than the ones found in \cite{Brout_Scolnic_2021} ($\Delta \mathbb{E}(R_V)=1.25\pm0.43$) and \cite{Popovic_2023} ($\Delta \mathbb{E}(R_V)=0.89\pm0.45$), who both attribute all of the environmental step to dust variations.

The only significant change is the $\sim 7\sigma$ difference in $\tau_A$ when splitting on local mass. This is in line with results from \cite{Ginolin_2025a}, who found that local mass, along with $d_\mathrm{DLR}$, had the biggest impact on dust attenuation, using the SALT colour distribution. Including a $d_\mathrm{DLR}$ dependency in the model would be an interesting avenue, especially in light of the results from \cite{Toy_2025}, who found a reduced mass step at high $d_\mathrm{DLR}$, i.e. in the outskirts of galaxies.

\subsection{Intrinsic steps}
\label{subsec:step_values}

We show in Table~\ref{tab:steps_dust} the estimates of the intrinsic steps, defined as $\Delta M_0^\mathrm{LM}-\Delta M_0^\mathrm{HM}$,  for the four environmental tracers.

\begin{table*}
 \caption{Environmental intrinsic step values ($\Delta M_0^\mathrm{LM}-\Delta M_0^\mathrm{HM}$) for the \citetalias{Grayling_2024} BayeSN model with modifications described in Sec.~\ref{subsec:step_method}, along with differences in mean $R_V$ and $\tau_A$ between low-mass/blue and high-mass/red host galaxies.}
 \label{tab:steps_dust}
 \begin{tabular}{lccc}
  \hline
  Step tracer & Intrinsic step (mag) & $\Delta\mathbb{E}(R_V)$ & $\Delta\tau_A$\\
   \hline
   $(g-z)_\mathrm{local}$ & $0.085\pm0.019$ ($4.5\sigma$)& $0.165\pm0.132$ ($1.2\sigma$) &  $-0.047\pm0.028$ ($1.7\sigma$)\\
   $(g-z)_\mathrm{global}$ & $0.072\pm0.019$ ($3.7\sigma$) & $0.155\pm0.126$ ($1.2\sigma$) & $-0.045\pm0.030$ ($1.5\sigma$)\\
   $\log(M_\star/M_\odot)_\mathrm{local}$ & $0.081\pm0.019$ ($4.3\sigma$) & $0.086\pm0.148$ ($0.6\sigma$) & $-0.203\pm0.030$ ($6.8\sigma$)\\
   $\log(M_\star/M_\odot)_\mathrm{global}$ & $0.103\pm0.018$ ($5.6\sigma$)& $0.146\pm0.139$ ($1.0\sigma$)& $-0.035\pm0.030$ ($1.2\sigma$) \\
  \hline
 \end{tabular}
\end{table*}

For every tracer, we detect a step at a $>3.5\sigma$ significance. This is a strong hint that the magnitude step in SNe~Ia is not fully due to dust, and has a intrinsic component. This is in line with results from \cite{Ginolin_2025a}, who found that the steps did not significantly vary with colour, contrary to what is expected for a dust-based model of the step, where blue (e.g. unaffected by dust) SNe have a smaller step than red (e.g. dusty) SNe.

The ZTF steps are slightly larger than the one estimated by \citetalias{Grayling_2024} from a compilation of Foundation, DES3Y and PS1MD SNe ($0.049\pm0.016$ mag, a $3.1\sigma$ significance). As the mean redshift of this sample is higher than the one of ZTF, this mismatch might be due to an evolution with redshift of the magnitude step. \citetalias{Grayling_2024} also used a volume-limited sample in their analysis, which diminishes the probability of selection effects explaining the larger step found in ZTF. 

There is a slight trend of higher $\tau_A$ and $R_V$ differences leading to a smaller step in Table~\ref{tab:steps_dust}, as the largest step (using global mass) is associated differences in $R_V$ and $\tau_A$ both at a $\sim 1 \sigma$ level. This is not surprising, as a larger variability in the extinction will lead to bigger differences in magnitudes, which then do not have to be compensated by the achromatic step.

The intrinsic steps found are smaller than the fiducial steps from \cite{Ginolin_2025b} using a SALT-based standardisation, which ranged from $\sim0.12$ mag to $\sim 0.17$ mag. This reduction might be due to the differing dust properties, which absorb part of the environmental dependencies of SN~Ia absolute magnitude. Retraining the model with an achromatic step and dust parameters common to all SNe gives slightly increased steps, with a local colour step of $0.101\pm0.014$ ($7.1\sigma$) and a global mass step of $0.118\pm0.015$ ($8.0\sigma$). However, a full retraining of the model including a step would be necessary to measure accurately the size of the step, so that all covariances between parameters are correctly accounted for. Indeed, as $\theta_1$ is strongly correlated with environment (see Fig.~\ref{fig:theta_distrib}), $W_1$ might absorb some of the environment-magnitude correlation, leading to an overall smaller step.

\section{Discussion}
\label{sec:discussion}

In this section, we discuss variations of our analysis, as well as possible extensions for future work.

\subsection{Impact of the training}
\label{subsec:training_impact}

\begin{table}
 \caption{Same as Table~\ref{tab:hubble_res_scatter} but for the \citetalias{Thorp_2021} training.}
 \label{tab:hubble_res_scatter_T21}
 \begin{tabular}{lcccc}
  \hline
  Scatter & \multicolumn{2}{c}{Full sample} & \multicolumn{2}{c}{Host-$z$ sample} \\
   & BayeSN & SALT & BayeSN & SALT\\
   \hline
   $\sigma_\mathrm{STD}$ & 0.235 & 0.230 & 0.214 & 0.213 \\
  $\sigma_\mathrm{nMAD}$ & 0.189 & 0.181 & 0.170 & 0.169 \\
  \hline
 \end{tabular}
\end{table}

\begin{table}
 \caption{Values of the magnitude steps (in mag) for the \citetalias{Thorp_2021} training, either computed after light curve fits ('a posteriori', same as Table~\ref{tab:bayesn_vs_salt_steps}) or inferred as a population parameter accounting for different dust properties ('intrinsic', same as Table~\ref{tab:steps_dust}).}
 \label{tab:steps_T21}
 \begin{tabular}{lcc}
  \hline
  Step tracer & A posteriori steps & Intrinsic steps \\
   \hline
   $(g-z)_\mathrm{local}$ & $0.093\pm0.011$ ($8.6\sigma$) & $0.073\pm0.019$ ($3.9\sigma$) \\
   $(g-z)_\mathrm{global}$ & $0.085\pm0.011$ ($7.9\sigma$) & $0.053\pm0.019$ ($2.8\sigma$)\\
   $\log(M_\star/M_\odot)_\mathrm{local}$ & $0.099\pm0.011$ ($9.3\sigma$) & $0.064\pm0.019$ ($3.5\sigma$) \\
   $\log(M_\star/M_\odot)_\mathrm{global}$ & $0.107\pm0.011$ ($10.0\sigma$) & $0.094\pm0.018$ ($5.3\sigma$)\\
  \hline
 \end{tabular}
\end{table}

To evaluate the impact of the use of the \citetalias{Grayling_2026b} training, we replicate our analysis with the training from \citetalias{Thorp_2021}. The biggest difference with our fiducial analysis is the scatter of the residuals, which are slightly larger than the SALT scatter for the full sample, and in line for the host-$z$ sample, as shown in Table~\ref{tab:hubble_res_scatter_T21}. Moreover, the steps, either for the a posteriori case or the the ones from the retraining are all compatible, as shown in Table~\ref{tab:steps_T21}. There is a slight outlier ($2.0\sigma$ difference between the \citetalias{Thorp_2021} and \citetalias{Grayling_2026b} trainings for the a posteriori step in local mass), but this is not unexpected considering the number of parameters we infer. We thus conclude that all of the results from Secs.~\ref{sec:lc_fits}-\ref{sec:step}, bar the Hubble residual scatter, are robust to a change of BayeSN training.

\subsection{Impact of the smooth step}
\label{subsec:smooth_step_impact}

\begin{table*}
 \caption{Same as Table~\ref{tab:steps_dust} but for a sharp step, as done in \citetalias{Grayling_2024}.}
 \label{tab:steps_dust_G24}
 \begin{tabular}{lccc}
  \hline
  Step tracer & Intrinsic step (mag) & $\Delta\mathbb{E}(R_V)$ & $\Delta\tau_A$\\
   \hline
   $(g-z)_\mathrm{local}$ & $0.072\pm0.019$ ($3.9\sigma$)& $0.165\pm0.122$ ($1.4\sigma$) &  $-0.040\pm0.029$ ($1.4\sigma$)\\
   $(g-z)_\mathrm{global}$ & $0.069\pm0.019$ ($3.6\sigma$) & $0.133\pm0.128$ ($1.0\sigma$) & $-0.043\pm0.029$ ($1.5\sigma$)\\
   $\log(M_\star/M_\odot)_\mathrm{local}$ & $0.075\pm0.020$ ($3.8\sigma$) & $0.070\pm0.135$ ($0.5\sigma$) & $-0.190\pm0.030$ ($6.4\sigma$)\\
   $\log(M_\star/M_\odot)_\mathrm{global}$ & $0.085\pm0.019$ ($4.4\sigma$)& $0.128\pm0.122$ ($1.1\sigma$)& $-0.030\pm0.030$ ($1.0\sigma$) \\
  \hline
 \end{tabular}
\end{table*}

To quantify the impact of using a smooth cut on environmental parameter, we run the analysis of Sec.~\ref{sec:step} but using a sharp cut, as was done in \citetalias{Grayling_2024}. 
In the regular SALT standardisation framework, this inclusion of a smooth step pushes $\gamma$ towards higher values, as it limits the impact of misspecified cases or SNe with large errors.
In the case of BayeSN, this would impact both the achromatic step and the dust distribution, leading to a better separation of the two. The resulting steps and dust parameters are computed in Table~\ref{tab:steps_dust_G24}. Implementing a smooth step pushed the step values higher, but within errorbars. This is in line with the behaviour seen in \cite{Ginolin_2025b}. The differences in dust properties, $\Delta \mathbb{E}(R_V)$ and $\Delta\tau_A$, were also pushed higher by the inclusion of the smooth step.

\subsection{Non-linearity of the $\theta_1$-Hubble residuals relation}
\label{subsec:broken_theta}

\begin{figure}
    \centering
    \includegraphics[width=1\columnwidth]{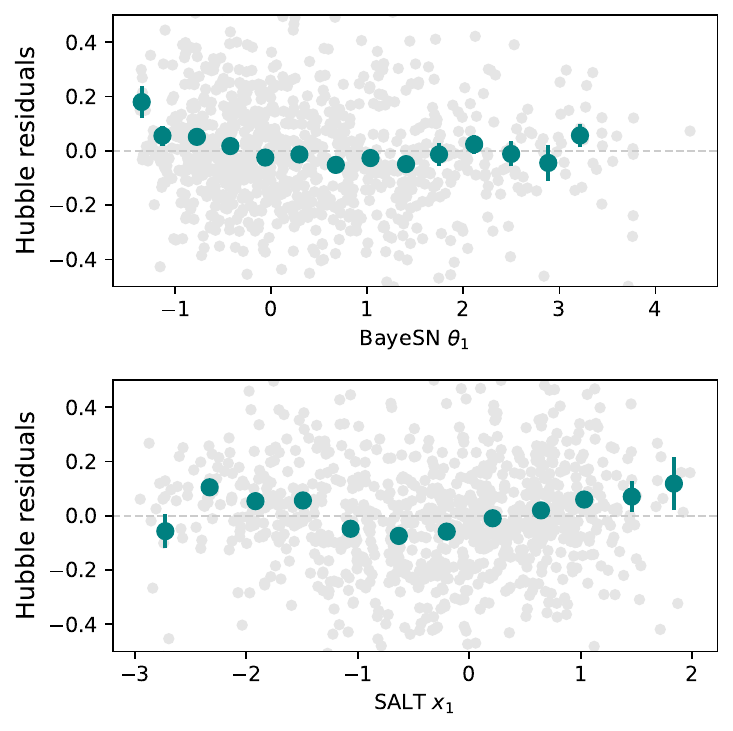}
    \caption{Hubble residuals, computed as described Sec.~\ref{subsec:hubble_res} (using a single $\alpha$), against BayeSN $\theta_1$ (top) and SALT $x_1$ (bottom). The grey points are individual SNe and the blue points are binned averages, with the errorbars denoting the error on the mean.}
    \label{fig:hubble_res_vs_theta}
\end{figure}

In \cite{Ginolin_2025b}, a strong ($\sim 13\sigma$) non linearity was found in the stretch-residuals relation (see also \citealt{Garnavich_2023, Larison_2024, Rubin_2026}). This translated into two different standardisation coefficients $\alpha$ for the two stretch regimes. In BayeSN, the surface $W_1$ encapsulate both the changes in the shape of the SED and the slower-brighter relation, whereas in SALT-based standardisation, those two effects are separated into an $M_1$ surface and an $\alpha$ coefficient, such that $\alpha\times M_1$ roughly corresponds to $W_1$. To investigate the existence of this non-linearity in BayeSN, we first look at any trend in the Hubble residuals-$\theta_1$ plane, as shown in the top of Fig.~\ref{fig:hubble_res_vs_theta}. As a point of reference, the same plot for SALT-standardised Hubble residuals is shown in the bottom panel of Fig.~\ref{fig:hubble_res_vs_theta}. There is not clear trend in the BayeSN Hubble residuals, contrary to the SALT ones. To quantify this effect, we refit the light curves but implementing a broken-$\alpha$ like standardisation in Eq.~\ref{eq:bayesn}. We thus replace $\theta_1^s\times W_1(t,\lambda)$ by:
\begin{equation}
    \begin{cases} \alpha_\mathrm{low}\theta_1^s\times W_1(t,\lambda)& \mbox{if }\theta_1^s \leq \theta_1^\mathrm{break} \\
    (\alpha_\mathrm{high}\theta_1^s + C)\times W_1(t,\lambda)& \mbox{if }\theta_1^s > \theta_1^\mathrm{break}\end{cases}.
    \label{eq:broken_w1}
\end{equation}
This adds three hyperparameters to the model, $\alpha_\mathrm{high}$, $\alpha_\mathrm{low}$ and $\theta_1^\mathrm{break}$. C is a constant that ensures continuity, and its value is fully determined by the set of ($\alpha_\mathrm{high}$, $\alpha_\mathrm{low}$, $\theta_1^\mathrm{break}$). We then retrain the model, keeping $W_0$, $W_1$ and $\mathbf{\Sigma_\epsilon}$, as well as the dust parameters $R_V$ and $\tau_A$, fixed to their model value. As $W_1$ is fixed, we have to add two coefficients, but if we were to retrain the full model, only one would be needed. 
\begin{figure}
    \centering
    \includegraphics[width=1\columnwidth]{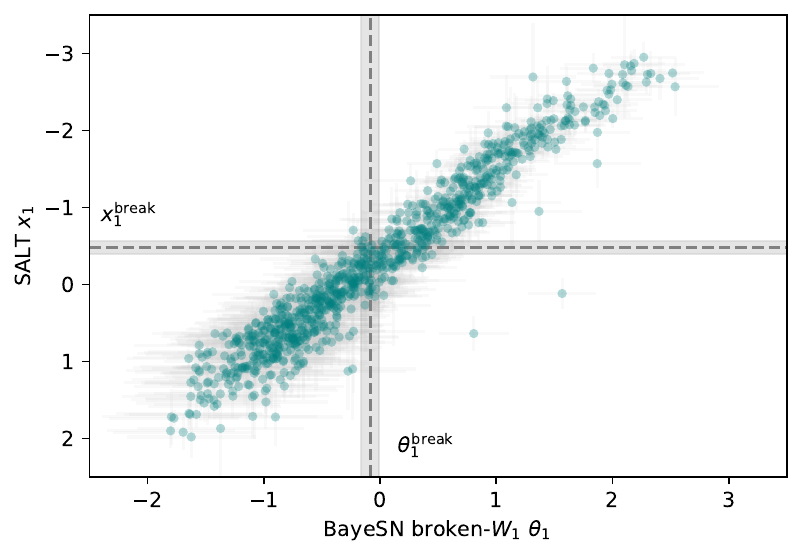}
    \caption{$\theta_1$ from the 'broken-$W_1$' model (see Sec.~\ref{subsec:broken_theta}) against SALT $x_1$. The shaded bands represent the fitted $\theta_1^\mathrm{break}$ and $x_1^\mathrm{break}$ and the corresponding $1\sigma$ errorbars.}
    \label{fig:broken_theta_vs_x1}
\end{figure}

\begin{figure}
    \centering
    \includegraphics[width=1\columnwidth]{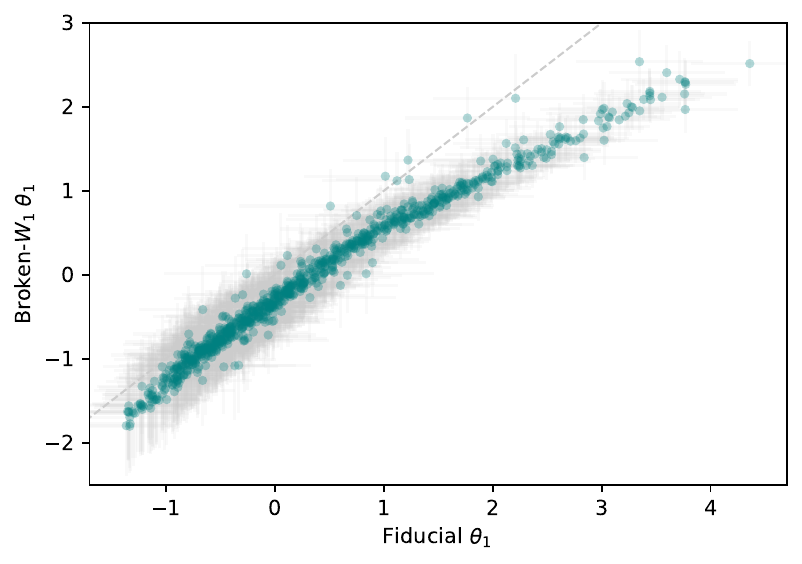}
    \caption{$\theta_1$ from the \citetalias{Grayling_2026b} BayeSN model (see Sec.~\ref{subsubsec:theta_distrib}) against $\theta_1$ in the modified 'broken-$W_1$' model, as investigated in Sec.~\ref{subsec:broken_theta}.}
    \label{fig:broken_theta_vs_fiducial}
\end{figure}

We find $\alpha_\mathrm{low}=0.616\pm0.030$ and $\alpha_\mathrm{high}=1.797\pm0.084$. This is in accordance with the results from \cite{Ginolin_2025b}. Indeed, low $\theta_1$, i.e. high stretch, SNe are more standard than high $\theta_1$ ones. Moreover, the difference in $\alpha$ corresponds to $\Delta\alpha=1.181\pm0.090$, as $13.2\sigma$ difference, in line with the $13.4\sigma$ level detected with SALT standardisation. We find $\theta_1^\mathrm{break}=-0.085\pm0.076$, roughly corresponding to the SALT $x_1^\mathrm{break}\sim -0.5$, as visible in Fig.~\ref{fig:broken_theta_vs_x1}.

However, in the BayeSN framework, refitting for those hyperparameters requires the joint fit of individual SN parameters, such as $\theta_1^s$. There is thus a degeneracy in the model, where the inclusion of $\alpha_\mathrm{high/low}$ can be counteracted with global shift of $\theta_1$ values for high/low $\theta_1$ SNe. This is visible in Fig.~\ref{fig:broken_theta_vs_fiducial}. While the low $\theta_1$ SNe seem relatively unchanged from the fiducial fit of Sec.~\ref{sec:lc_fits}, the high $\theta_1$ SNe see their $\theta_1$ values being shifted lower by around $\sim1$. $A_V$ are largely unaffected. Looking at the comparison between SALT $x_1$ and $\theta_1$ when including the 'broken-$W_1$' relation in Fig.~\ref{fig:broken_theta_vs_x1}, they are now much more linearly correlated, without the curved tail visible at high $\theta_1$ in Fig.~\ref{fig:x1_vs_theta}.

A tentative explanation of this behaviour is that in the fiducial analysis of Sec.~\ref{sec:lc_fits}, the BayeSN model is trying to grasp the non-linear relation between BayeSN $\theta_1$ and Hubble residuals by systematically increasing $\theta_1$ for the high $\theta_1$ SNe, which boosts their effective $\alpha$. This is because BayeSN encodes the standardisation in its model, as all SNe are assumed to share the same absolute magnitude $M_0$. When giving the model an extra degree of freedom to account for the 'broken-$W_1$' standardisation, this boost is encapsulated in the $\alpha_\mathrm{high/low}$ parameters, and the $\theta_1$ parameters revert to SALT-like values.

A principled way to account for these different standardisation relations for high- and low-stretch SNe would be to use a mixture model to account for two stretch populations, as was done in \cite{Rubin_2026, Wojtak_2026}. This is however outside the scope of this work.

When computing an a posteriori step with this retraining, as done in Sec.~\ref{subsec:steps_aposteriori} and reported in Table~\ref{tab:bayesn_vs_salt_steps}, we find a global mass step of $0.097\pm0.009$ mag, a $10.1\sigma$ significance. This is a negligible change compared to our fiducial analysis ($0.103\pm0.010$, a $10.1\sigma$ significance). We conclude that the (ad-hoc) inclusion of a 'broken-$W_1$' standardisation does not impact the environmental step.

\subsection{Intrinsic colour variations between SNe from different environments}
\label{subsec:intrinsic_colour_step}

\begin{table*}
 \caption{Same as Table~\ref{tab:dust_params}, but for the model described in Sec.~\ref{subsec:intrinsic_colour_step}, where $\boldsymbol{\Sigma}_\epsilon$ varies with environment.}
 \label{tab:dust_params_intrinsic_colour}
 \begin{tabular}{lccccc}
  \hline
  Sample & $\tau_A$ & $\mu_R$ & $\sigma_R$ & $\mathbb{E}(R_V)$ & $\sqrt{\mathrm{Var}(R_V)}$ \\
  \hline
   \hline
   $\log(M_\star/M_\odot)_\mathrm{global} > 10$ & $0.446\pm0.024$ & $2.349\pm0.083$ & $0.448\pm0.071$ & $2.359\pm0.081$ & $0.437\pm0.062$\\
  $\log(M_\star/M_\odot)_\mathrm{global} < 10$ & $0.365\pm0.023$ & $2.285\pm0.134$ & $0.664\pm0.119$ & $2.370\pm0.102$ & $0.591\pm0.077$\\
  \hline
 \end{tabular}
\end{table*}

In Sec.~\ref{sec:step}, we assumed a common intrinsic colour distribution for SNe in high- and low-mass hosts. We relax this assumption in this section and retrain our model with an additional split in $\boldsymbol{\Sigma}_\epsilon$. Using global mass as the environmental tracer, we find an intrinsic step of $0.111\pm0.020$ (a $5.1\sigma$ significance). This is compatible with Sec.~\ref{sec:step}, where we found an intrinsic step in global mass of $0.103\pm0.018$ ($5.6\sigma$). We see small changes in the difference of dust properties between high- and low-mass SNe, with $\Delta\mathbb{E}(R_V)=0.012\pm0.134$ (0.1$\sigma$) and $\Delta\tau_A=-0.080\pm0.033$ (2.4$\sigma$), but they remain insignificant. 
Interpreting the effect of $\boldsymbol{\Sigma}_\epsilon$ is more challenging. We simulate light curves of 10,000 SNe by sampling from $\boldsymbol{\Sigma}_\epsilon^\mathrm{HM/LM}$ while keeping $\theta_1=0$, $A_V=0$, and we calculate $(B-V)$ colours at peak, as this is a close estimate of a SALT-like intrinsic colour. As expected, the mean $(B-V)$ colours are similar for the base model ($\mu_{(B-V)}=-0.140$), the high-mass model ($\mu_{(B-V)}^{HM}=-0.141$) and the low-mass model ($\mu_{(B-V)}^{LM}=-0.142$). The low-mass model shows a slightly lower scatter ($\sigma_{(B-V)}^{LM}=0.031$) than the high-mass model ($\sigma_{(B-V)}^{LM}=0.031$) and the nominal one ($\sigma_{(B-V)}=0.041$). These scatters are comparable to the ones found in Sec.~\ref{subsec:salt_comparison}. We note that in redder bands, the intrinsic colour distributions are essentially the same across environments.
This additional degree of freedom did not change our results: the intrinsic step is detected at more than $5\sigma$, and the dust properties do not vary significantly across environment.

\subsection{Intrinsic SED variability between SNe from different environments}
\label{subsec:sed_step}

On top of fitting for an achromatic step, as we do in Sec.~\ref{sec:step}, \citetalias{Grayling_2024} also investigated intrinsic SED differences between SNe in different environments by fitting for different $W_0$ surfaces for SNe in high- and low-mass hosts. Their main finding was a strong difference in PS1 $i$-band and PS1 $r$-$i$ colour at $\sim 20$ days after peak. We here reproduce this analysis, adding the smooth transition between the two SN subpopulations as described in Sec.~\ref{subsec:step_method}. For this analysis variation, we only split on global host mass. We use a slightly different post-processing than in \citetalias{Grayling_2024}, that we describe in Appendix \ref{app:sed_split}. This post-processing is needed to account for degeneracies between global shifts of $\theta_1$ and changes in $W_0$.

\begin{figure*}
    \centering
    \includegraphics[width=2\columnwidth]{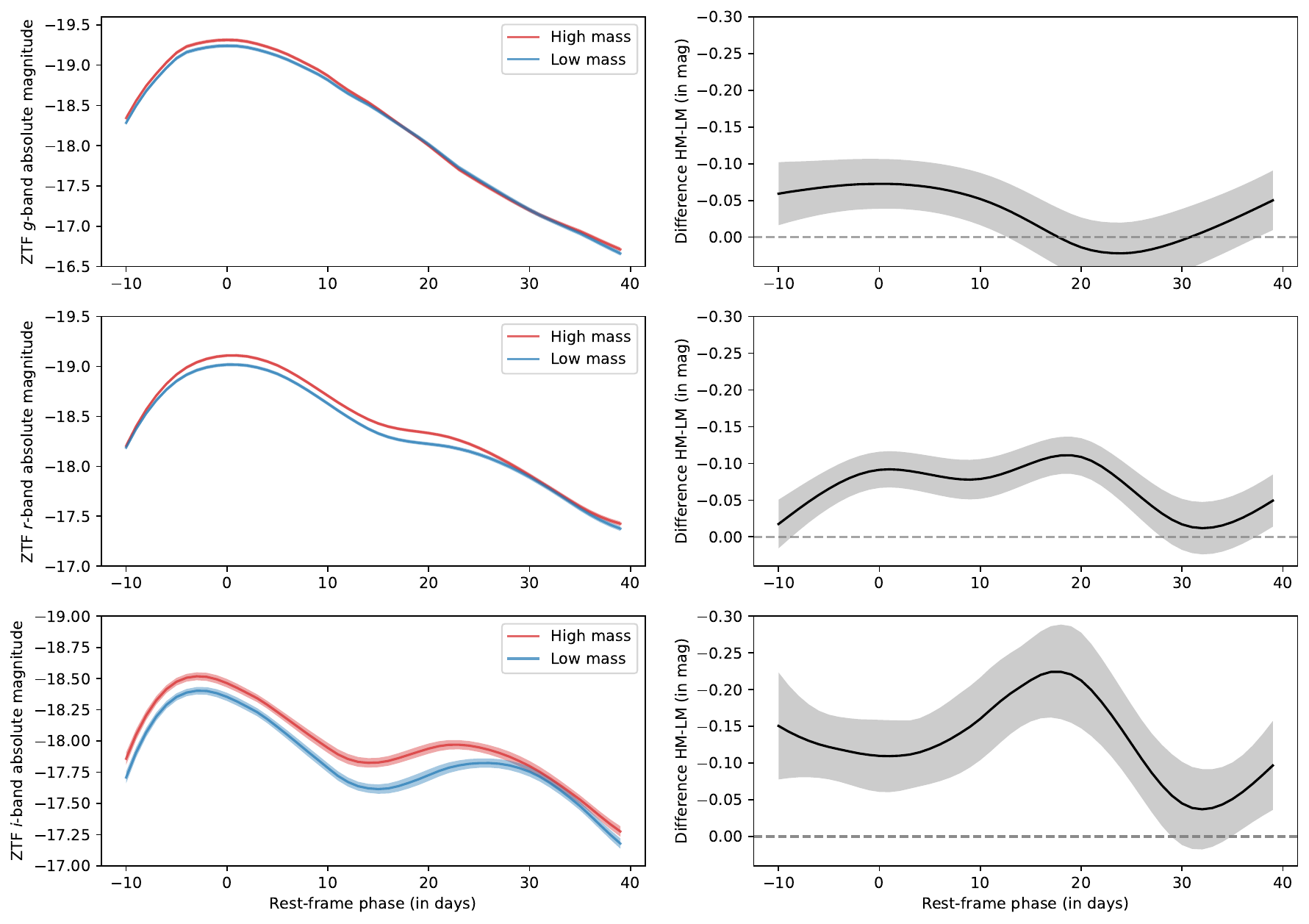}
    \caption{\textit{Left:} Light curve of a $A_V=0$, $\theta_1=0$, $\epsilon=0$ SN in a high-mass (red) and low-mass (blue) host galaxy, in ZTF $g$, $r$ and $i$ bands. The model has been trained using the method in Sec.~\ref{subsec:sed_step}, and post-processed according to Appendix~\ref{app:sed_split}. \textit{Right:} Difference between the high- and low-mass models.}
    \label{fig:sed_split_lcs}
\end{figure*}

Looking at Fig.~\ref{fig:sed_split_lcs}, we see that the intrinsic SED differences between low- and high-mass SNe are stronger in $i$-band, especially at $\sim 20$ days after peak. We measure a $0.073\pm0.034$ mag ($2.1\sigma$) step in peak $g$-band, a $0.091\pm0.025$ mag ($3.7\sigma$) step in peak $r$-band and a $0.110\pm0.049$ mag ($2.2\sigma$) step in peak $i$-band. While theses step are smaller than the step in $i$-band at 20 days after peak ($0.212\pm0.065$ mag, $3.3\sigma$), the larger error in $i$-band, due to the poorer coverage (only half of the sample has $i$-band light curves, see Sec.~\ref{subsec:full_training}), lowers its significance. These results confirm an intrinsic variability in SED between SNe in high- and low-mass hosts, which fluctuates with wavelength and time, as found in \citetalias{Grayling_2024}. When extending the SALT model to add an extra parameter $x_2$, \cite{Kenworthy_2025} found that this parameter allowed for phase-dependent variations of colours curves, linked with the height of the $i$-band secondary maxima. \cite{Deckers_2025} also found than standardising SNe with the strength of the light curve secondary maximum gave a better standardisation than using the SALT stretch $x_1$, which could be due to the fact that it encapsulates some of the magnitude step. Further constraining this variability, with large samples and wide wavelength coverage, might be an interesting avenue to understand the physical origin of the step.

\subsection{Full retraining}
\label{subsec:full_training}

The current version of the ZTF SN~Ia DR2 is not yet ready for cosmology. In particular, \cite{Lacroix_2026} find a 90 mmag calibration uncertainty in magnitude, along with a 10 mmag shift in colour. Moreover, 52\% of the SNe in the sample used in this analysis currently have detections in only two bands ($g$ and $r$), partly due to the lack of baseline measurements.

These issues are currently preventing the use of the DR2 as a training sample for BayeSN. Having sufficient wavelength coverage is particularly important to ensure sufficient constraints when inferring both intrinsic SED (e.g. $W_0$ and $W_1$) and dust properties ($\mu_R$, $\sigma_R$ and $\tau_A$).

Nonetheless, these issues should be fixed with the upcoming release of ZTF scene-modelling light curves. This data will provide an interesting test case for training BayeSN, due to its large number of SNe, its homogeneity and its limited selection bias.

\section{Conclusion}
\label{sec:conclusion}

In this paper, we fit the ZTF SN~Ia DR2 volume-limited sample with BayeSN. BayeSN is a hierarchical Bayesian model of SNe~Ia at the SED level, which explicitly models the dust impact on SN~Ia light curves. Our conclusions are the following:
\begin{itemize}
    \item Using the new \citetalias{Grayling_2026b} BayeSN training, we find a slightly reduced Hubble residuals scatter ($\sigma_\mathrm{nMAD}=0.177$) than when using SALT ($\sigma_\mathrm{nMAD}=0.181$). In comparison, the fiducial BayeSN model \citetalias{Thorp_2021} gives $\sigma_\mathrm{nMAD}=0.189$.
    \item We find a scatter in SALT $c$ for SNe with $A_V=0$, hinting at an intrinsic SN colour scatter. Fitting SNe with low $A_V$ ($A_V<0.1$), we find a characteristic $c$ scatter of $\sim 0.03$, in line with the value from \cite{Ginolin_2025a}. 
    \item We find a non-linearity between BayeSN $\theta_1$ and SALT $x_1$. This might be due to an attempt of the BayeSN model to account for the strong non-linearity of the stretch-residuals relation (or broken-$\alpha$) found in \cite{Ginolin_2025b}.
    \item We find a $0.103\pm0.010$ mag ($10.1\sigma$) global mass step and a $0.086\pm0.010$ ($8.3\sigma$) local colour step when using the Hubble residuals from the BayeSN light curve fits, in accordance with the a posteriori SALT steps from \cite{Ginolin_2025b}. This means that the large steps seen in ZTF were not due to SALT or to the standardisation process.
    \item Accounting for different dust properties in high- and low-mass galaxies, we find a global mass step of $0.103\pm0.018$ ($5.6\sigma$) and a local colour step of $0.085\pm0.019$ ($4.5\sigma$). The difference in mean $R_V$ between low- and high-mass galaxies is non-significant, respectively at $\Delta\mathbb{E}(R_V)=0.146\pm0.139$ ($1.0\sigma$) for global mass and $\Delta\mathbb{E}(R_V)=0.165\pm0.132$ ($1.2\sigma$) for local colour. We thus conclude that the environmental step is at least partially intrinsic.
\end{itemize}

Upcoming low-redshift SN~Ia surveys, such as TITAN \citep{TITAN}, DEBASS \citep{DEBASS} and YSE \citep{Jones_2021}, will be valuable to understand the origin of the large step in the ZTF sample. Detailed studies of the host environment of SNe Ia, using either multi-bands host galaxy measurements \citep[e.g. ][]{Ramaiya_2025} or simulations \citep[e.g. ][]{Wiseman_2022}, will also bring insight into the origin of the step. Nevertheless, our results reaffirm the strength of this effect, and thus the need to correctly account for it in cosmological analyses to guarantee unbiased cosmological parameter inference.

\section*{Acknowledgements}
\addcontentsline{toc}{section}{Acknowledgements}

Supernova and astrostatistics research at Cambridge University is supported in part by the European Union’s Horizon 2020 research and innovation programme under European Research Council Grant Agreement No 101002652 (BayeSN; PI K. Mandel).
LK acknowledges support for an Early Career Fellowship from the Leverhulme Trust through grant ECF-2024-054 and the Isaac Newton Trust through grant 24.08(w).


\section*{Data Availability}

The data used in this analysis can be downloaded from \href{https://ztfcosmo.in2p3.fr}{ztfcosmo.in2p3.fr}. The BayeSN code is available at \href{https://github.com/bayesn}{\texttt{github.com/bayesn}}, its version used for this article is available at \href{https://github.com/mginolin/bayesn}{\texttt{github.com/mginolin/bayesn}}, and the \texttt{standax} code is available at \href{https://github.com/mginolin/standax}{\texttt{github.com/mginolin/standax}}.


\bibliographystyle{mnras}
\bibliography{mnras} 

\newpage
\appendix
\section{Post-processing of the model for the intrinsic SED variability}
\label{app:sed_split}

Looking at Eq.~\ref{eq:bayesn}, there exists a degeneracy between the definition of $\theta_1$ and $W_0$. Indeed, a global shift of $\theta_1 \rightarrow \theta_1 + A$ combined with a shift of $W_0 \rightarrow W_0 - A \times W_1$ leaves the model unchanged. This degeneracy is partially broken by the use of a prior on $\theta_1$, but it means that the definition of $\theta_1$ is tied to the data the model was trained on. This comes as an issue when training separately on a high and low-mass subsample, as done in Sec.~\ref{subsec:sed_step}, where both subsamples will end up with a $\theta_1$ distribution mean around 0, whereas their light curve widths are on average different (see Fig.~\ref{fig:theta_distrib}). As this $\theta_1$ shift then affects $W_0$, some post-processing of the output parameters is needed in order to be able to compare the high and low-mass SEDs. 
The post-processing used in \citetalias{Grayling_2024} anchored $\theta_1$ to the decline rate in PS1-$g$-band $\Delta m_{15}$, so that SNe in a high and low-mass galaxy with $\theta_1=0$, $A_V=0$, $\epsilon=0$ had the same $\Delta m_{15}$. However, this post-processing relied on two hypotheses (independence of $\frac{\partial\Delta m_{15}}{\partial\theta_1}$ on $\Delta m_{15}$, independence of $\Delta m_{15}$ on $A_V$), valid when using the \citetalias{Thorp_2021} model but no longer verified with the \citetalias{Grayling_2026b} model. This may be due to the wider range of colour and light curve widths spanned by the \citetalias{Grayling_2026b} training sample, probing parameter spaces unexplored before, as well as the non-linearity of the light-curve width-magnitude relation, that the currently linear BayeSN model does not account for. We thus modify the post-processing so that the $\theta_1$ distributions are anchored to the ones previously obtained when fitting for an achromatic intrinsic step, as in Sec.~\ref{sec:step}. We compute the low- and high-mass $\theta_1$ shifts as:
\begin{equation}
    \Delta \theta_1^\mathrm{LM} = \frac{\sum_s p^s (\theta_{1\mathrm{,~SED}}^{s} - \theta_{1\mathrm{,~ref}}^{s})}{\sum_s p^s },
\end{equation}
\begin{equation}
    \Delta \theta_1^\mathrm{HM} = \frac{\sum_s (1-p^s) (\theta_{1\mathrm{,~SED}}^{s} - \theta_{1\mathrm{,~ref}}^{s})}{\sum_s (1-p^s) },
\end{equation}
where $\theta_{1, \mathrm{ref}}$ is the reference $\theta_1$ value when fitting for an achromatic step, and $p^s$ is the probability of a given SN to have a low-mass host (see Sec.~\ref{subsec:step_method}). We then correct the $W_0$ surfaces using $W_0^\mathrm{LM/HM} \rightarrow W_0^\mathrm{LM/HM} - \Delta \theta_1^\mathrm{LM/HM} W_1$.

\bsp	
\label{lastpage}
\end{document}